\documentclass[prl,twocolumn]{revtex4}

\usepackage{amsmath}
\usepackage{amsthm}
\usepackage{graphicx}

\theoremstyle{remark}

\begin{document}

\newtheorem{lem}{Lemma}
\newtheorem*{prop}{Theorem}

\title{Internal entanglement and external correlations 
of any form limit each other}
\author{S. Camalet}
\affiliation{Sorbonne Universit\'e, CNRS, Laboratoire de Physique
 Th\'eorique de la Mati\`ere Condens\'ee, LPTMC, F-75005, 
Paris, France}

\begin{abstract}
We show a relation between entanglement and correlations 
of any form. The internal entanglement of a bipartite system, 
and its correlations with another system, limit each other. 
A measure of correlations, of any nature, cannot increase 
under local operations. Examples are the entanglement 
monotones, the mutual information, that quantifies total 
correlations, and the Henderson-Vedral measure of classical 
correlations. External correlations, evaluated by such 
a measure, set a tight upper bound on the internal 
entanglement that decreases as they increase, and so does 
quantum discord.
\end{abstract} 

\maketitle

Quantum entanglement is a useful resource for many tasks, such 
as cryptographic key distribution \cite{E}, state teleportation 
\cite{BBCJPW}, or clock synchronization \cite{JADW}, to cite just 
a few. In more precise terms, it is a quantum resource that cannot 
be generated by local operations and classical communication 
\cite{HHHH,V,PV}. The corresponding so-called free states, for 
which the resource vanishes, are the separable states, that are 
the mixtures of product states. Accordingly, entanglement is 
quantified by measures, termed entanglement monotones, which 
are non-negative functions of quantum states that vanish for 
separable states, and are nonincreasing under state 
transformations involving only local operations and classical 
communication.

Two real systems, whose entanglement is of interest, are 
never completely isolated from the surroundings. Consequently, 
a third system, which cannot be fully controlled, always comes 
into play. Using Hamiltonian models describing the influence of 
more or less realistic environments, different dynamic 
behaviours of the entanglement have been found, depending, 
for example, on whether the environment is in thermal 
equilibrium or not. For instance, an initial entanglement can 
vanish in finite time \cite{YE}, or, on the contrary, 
entanglement can develop transiently \cite{B,FT}, or even be 
steady \cite{CKS,EPJB,PRAo}. 

The impact of the surroundings on entanglement can also 
be approached by studying how entanglement is distributed 
between three systems in an arbitrary state. The amounts 
of entanglement between one of them and each of the two 
other ones constrain each other. This behaviour, known as 
entanglement monogamy, has first been shown for three 
two-level systems, and expressed as an inequality involving 
a particular entanglement monotone \cite{CKW}. This inequality 
does not hold in general for familiar monotones such as 
the entanglement of formation, or the regularized relative 
entropy of entanglement. For these two measures, inequalities 
involving Hilbert space dimensions explicitly must be considered 
\cite{LDHPAW}. Relations have also been found between 
the amounts of entanglement for the three bipartitions of 
a tripartite system \cite{ZF}.

Recently, another restriction on the distribution of entanglement 
between three systems has been shown \cite{PRL}. It is better 
understood by considering a finite-dimensional 
bipartite system, say $A$, and any other system, say $B$, 
which can be seen as the environment of $A$. It has been 
found that the internal entanglement, between the two 
subsystems of $A$, and the external entanglement, 
between $A$ and $B$, limit each other. This relation is 
expressed by an inequality involving entanglement monotones 
and the Hilbert space dimensions of the subsystems of $A$. 
One may wonder whether this is a specific property of 
entanglement, or whether a similar relation exists between 
internal entanglement and external correlations of any kind.

In this Letter, we address this issue by using measures of 
external correlations, which we term correlation monotones. 
Such a measure $C$ is a non-negative function of the state 
$\rho$ shared by $A$ and $B$, that vanishes for product 
states, and is nonincreasing under local operations, which 
do not affect either $A$ or $B$. These are basic requirements 
for a measure of correlations, since correlations, whatever 
their nature, cannot increase when $A$ and $B$ evolve 
independently. Our main result relies essentially on them. 
To be more specific, our derivation does not require that 
$C$ is a strict correlation monotone, but only that it is 
invariant under unitary local operations, and nonincreasing 
under operations performed on $B$. 
Examples of correlation monotones are the entanglement 
monotones, the mutual information, commonly used to 
quantify total correlations, and the Henderson-Vedral (HV) 
measure of classical correlations \cite{HV}. Quantum discords, 
on the other hand, are not correlation monotones \cite{DVB}. 
However, the original quantum discord \cite{OZ} as measured 
by $A$, satisfies the above mentioned properties 
\cite{SKB,MBCPV}, and so our approach applies to it.  

\begin{figure}
\centering
\includegraphics[width=0.45\textwidth]{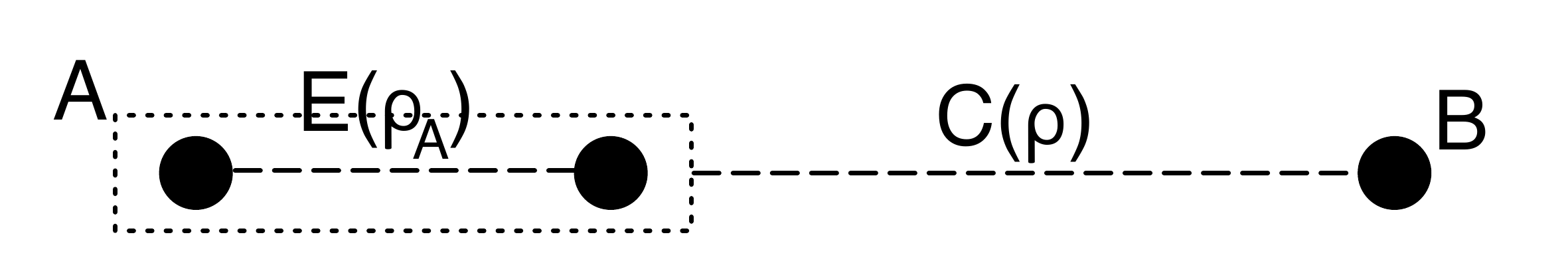}
\caption{Schematic representation of the systems and correlations 
considered.}
\label{S}
\end{figure}

\begin{figure}
\centering
\includegraphics[width=0.45\textwidth]{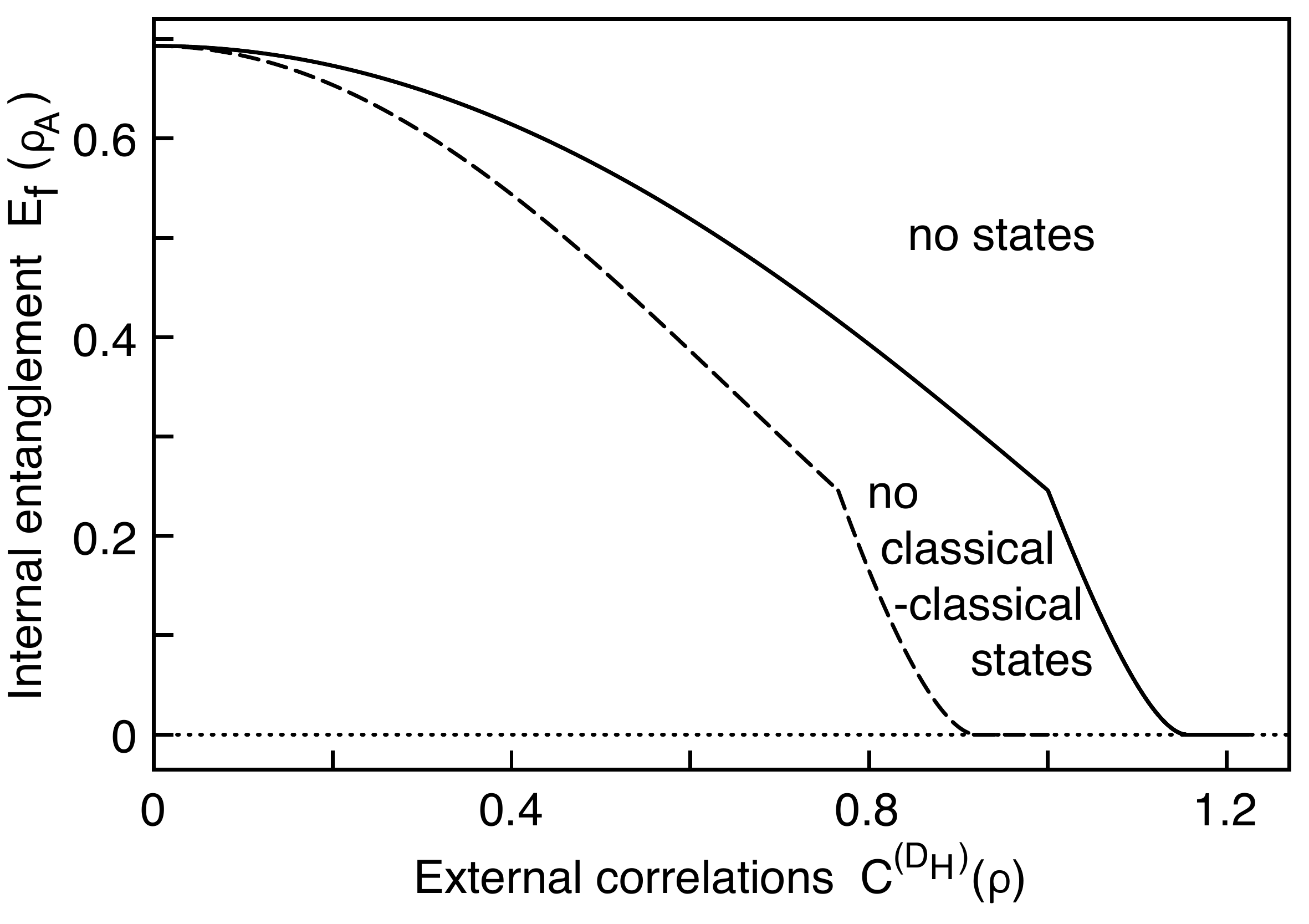}
\caption{Maximum internal entanglement as a function of 
external correlations, for a system $A$ consisting of two 
two-level systems, the entanglement of formation $E_f$, and 
the measure of total correlations $C^{(D_H)}$ (solid line). 
The maximum entanglement $E_f(\rho_A)$ for 
classical-classical states $\rho$ is given by the dashed line. 
This line is also the maximum value of $E_f(\rho_A)$ as 
a function of $C^{(D_B)}(\rho)$ for all states $\rho$.}
\label{Fig}
\end{figure}

We show in the following that, for an arbitrary finite-dimensional 
bipartite system $A$ and any system $B$, under an assumption 
of continuity usually fulfilled, $C(\rho)$ and the internal 
entanglement of $A$ are related to each other. More precisely, 
$C(\rho)$ determines a tight upper bound on $E(\rho_A)$, 
where $\rho_A$ is the reduced density operator for $A$, and 
$E$ is any convex entanglement monotone, that decreases as 
$C(\rho)$ increases, see Figs. \ref{S} and \ref{Fig}. As we will 
see, for familiar correlation monotones, this bound vanishes 
when $C(\rho)$ equals to its maximum value, set by the Hilbert 
space dimension of $A$. Moreover, since our result holds when 
$C$ is the HV measure, it implies that, even when the external 
correlations are purely classical, they have a detrimental 
influence on internal entanglement.

In the following, 
$\boldsymbol \lambda (\omega)$ refers to the vector made up of 
the nonzero eigenvalues of the quantum state $\omega$, in 
decreasing order. It is a probability vector, i.e., its components 
are positive and sum to unity. If $\omega$ is a density operator 
on the Hilbert space ${\cal H}_d$ of dimension $d$, 
$\boldsymbol \lambda (\omega)$ belongs to the set ${\cal E}_d$ 
of probability vectors of no more than $d$ components. We call 
entropy any non-negative function of the probability vectors 
$\boldsymbol p$, which is nondecreasing with disorder, in the sense 
of majorization \cite{MOA}, and vanishes for $\boldsymbol p=1$ 
\cite{K}. Any entropy has a largest value on ${\cal E}_d$, 
reached for the equally distributed vector $(1/d,\ldots,1/d)$, which 
is majorized by any $\boldsymbol p \in {\cal E}_d$, and possibly 
also for other vectors.

To derive our main result, we use the following three Lemmas. 
The proofs of the first and third are given in the Supplemental 
Material \cite{SM}. The second is proved in Ref.\cite{PRL}.
\begin{lem}\label{lem1}
For any correlation monotone $C$, there is a function $f$ of 
the probability vectors with $f(1)=0$, such that, for 
any global state $\rho$,
\begin{equation}
C(\rho) \le f[\boldsymbol \lambda (\rho_A)] , \label{ESloc}
\end{equation}
with equality when $\rho$ is pure.
\end{lem}
We denote by $c_d$ the supremum of $f$ on ${\cal E}_d$. 
Due to eq.\eqref{ESloc}, $C(\rho)$ cannot exceed $c_d$ 
when the Hilbert space of $A$ is ${\cal H}_d$. When $C$ 
is an entanglement monotone, $f$ is necessarily an entropy 
\cite{PRA2}. It is the Shannon entropy $h$ for many familiar 
entanglement monotones and for the HV measure 
\cite{HHHH,BBPS,VP,HV,ZWF}. For robustness and negativity, 
$f$ is a function of the R\'enyi entropy \cite{VT,ViWe,HN,S}. 
From the Araki-Lieb inequality 
$S(\rho) \ge |S(\rho_B)-S(\rho_A)|$, where $S$ is the von 
Neumann entropy \cite{AL}, it follows that $f=2h$ for the mutual 
information $S(\rho_A)+S(\rho_B)-S(\rho)$. As mentioned in 
the introduction, the quantum discord as measured by $A$, though 
not a correlation monotone, has the required properties to satisfy 
Lemma \ref{lem1}, see the proof. The corresponding function $f$ 
is $h$ \cite{MBCPV}. When $f$ is an entropy, $C$ coincides with 
an entanglement monotone for pure states \cite{N,HHHH}. For 
all the correlation monotones mentioned above, $f$ equals 
to $c_d$ for $(1/d,\ldots,1/d)$, and for no other vector of 
${\cal E}_d$. This means that, on the set of the pure states 
$|\psi \rangle$ of ${\cal H}_d \otimes {\cal H}_{d'}$, where 
$d' \ge d$, the maximally entangled states are the only ones 
for which $C(|\psi \rangle \langle \psi|)$ is maximum. 
\begin{lem}\label{lem2}
For any convex entanglement monotone $E$, and integers 
$d_1 \ge 2$ and $d_2 \ge d_1$, there are a positive number 
$e_{d_1}$ and an entropy $s_{d_1,d_2}$ such that the states 
$\rho_A$ on ${\cal H}_{d_1} \otimes {\cal H}_{d_2}$ satisfy
\begin{equation}
E(\rho_A) \le e_{d_1}
-s_{d_1,d_2}[\boldsymbol \lambda (\rho_A)] , \label{ES}
\end{equation}
and such that, for any 
$\boldsymbol p \in {\cal E}_{d_1\times d_2}$ and $\eta > 0$, 
there is $\rho_A$ for which 
$\boldsymbol \lambda (\rho_A)=\boldsymbol p$ and 
$e_{d_1}-s_{d_1,d_2}(\boldsymbol p)-E(\rho_A)<\eta$. 
\end{lem}
This Lemma expresses quantitatively how the mixedness of 
a quantum state limits its amount of entanglement \cite{ZHSL}. 
In Ref.\cite{PRL}, $e_{d_1}$ is obtained as the largest value 
of $E(\rho_A)$ for pure states $\rho_A$. Thus, it depends only 
on $d_1$ \cite{PV}. Inequality \eqref{ES} shows that it is 
the maximum of $E(\rho_A)$ on the set of all the density 
operators $\rho_A$ on 
${\cal H}_{d_1} \otimes {\cal H}_{d_2}$. Contrary to $e_{d_1}$, 
the entropy $s_{d_1,d_2}$ can depend on both $d_1$ and $d_2$, 
see the Supplemental Material. 
\begin{lem}\label{lem3}
For any positive integer $d$, entropy $s$, and nonnegative 
continuous function $f$ of the probability vectors with $f(1)=0$, 
there is a nondecreasing function $g_d$ on $I=[0,c_d]$, 
where $c_d$ is the maximum of $f$ on ${\cal E}_d$, 
such that $g_d(0)=0$, $g_d \circ f \le s$ on ${\cal E}_d$, and, 
for any $x \in I$ and $\eta>0$, there is 
$\boldsymbol p \in {\cal E}_d$ for which $f(\boldsymbol p)=x$ 
and $s(\boldsymbol p)-g_d(x)< \eta$.

If $f(1/d, \ldots, 1/d)=c_d$ and $f(\boldsymbol p)<c_d$ 
for any other $\boldsymbol p \in {\cal E}_d$, then 
$g_d(c_d)=s(1/d, \ldots, 1/d)$. 
\end{lem}
Using this Lemma with the function $f$ given by Lemma \ref{lem1}, 
and the entropy $s_{d_1,d_2}$ given by Lemma \ref{lem2}, 
and defining $\xi_{d_1,d_2}=e_{d_1}-g_d$, with $d=d_1d_2$, 
we have the following result.
\begin{prop}\label{prop1}
Let ${\cal H}_{d_1} \otimes {\cal H}_{d_2}$, with $d_2 \ge d_1$, 
be the Hilbert space of system $A$, and $d=d_1d_2$. 

For a convex entanglement monotone $E$, and a correlation 
monotone $C$ such that $f$ is continuous, $C(\rho)$ and 
$E(\rho_A)$ obey, for any global state $\rho$, 
\begin{equation}
E(\rho_A)  \le \xi_{d_1,d_2}[C(\rho)]  , \label{mineq}
\end{equation}
where $\xi_{d_1,d_2}$ is a nonincreasing function on $[0,c_d]$ 
with $\xi_{d_1,d_2}(0)=e_{d_1}$. 
For any amount of correlations $x \le c_d$, there are states $\rho$ 
such that $C(\rho)=x$ and the two sides of inequality 
\eqref{mineq} are as close to each other as we wish. 

If $f(1/d, \ldots, 1/d)=c_d$ and $f(\boldsymbol p)<c_d$ 
for any other $\boldsymbol p \in {\cal E}_d$, then 
$\xi_{d_1,d_2}(c_d)=0$. 
\end{prop}
Inequality \eqref{mineq} can be rewritten, in a more familiar form, 
as $E(A_1:A_2)  \le \xi_{d_1,d_2}[C(A_1A_2:B)]$, where $A_1$ 
and $A_2$ are the two subsystems of $A$, see Fig.\ref{S} 
\cite{PRL}. For any $x \in [0,c_d]$ and small $\eta$, Lemmas 
\ref{lem2} and \ref{lem3} ensure that there is a local state 
$\rho_A$ such that $\xi_{d_1,d_2}(x)-E(\rho_A)<\eta$ and 
$f[\boldsymbol \lambda (\rho_A)]=x$. Due to Lemma \ref {lem1}, 
all the pure states $\rho$ for which the reduced density operator 
for $A$ is $\rho_A$, are such that $C(\rho)=x$. For such global 
states $\rho$, $E(\rho_A) \simeq \xi_{d_1,d_2}[C(\rho)]$, 
and an increase of the correlations between $A$ an $B$ means 
a reduction of the internal entanglement of $A$, and reciprocally. 
In general, the external correlations and the local entanglement 
limit each other, see Fig.\ref{Fig}. For any amount of correlations 
$x \le c_d$, there is no state $\rho$ such that $C(\rho)=x$ and 
$E(\rho_A)$ exceeds $\xi_{d_1,d_2}(x)$. Similarly, for any amount 
of entanglement $y \le e_{d_1}$, there is no state $\rho$ such that 
$E(\rho_A)=y$ and $C(\rho)$ is larger than the bound given by 
eq.\eqref{mineq}. On the contrary, there are no positive lower 
bounds for $E(\rho_A)$, for a given $C(\rho)$, and for $C(\rho)$, 
for a given $E(\rho_A)$, whatever are the monotones $E$ and $C$ 
\cite{fn2}. 

For more than two systems, say $A$, $B_1$, $B_2$, \ldots, 
different bounds on the entanglement $E(\rho_A)$ can be 
obtained via eq.\eqref{mineq}, depending on which systems 
$B_n$ are taken into account. Let us first observe that only 
the systems sharing a state with genuine multipartite 
correlations matter \cite{BGHHH}. Indeed, if the global state 
is of the form $\rho=\tilde \rho \otimes \hat \rho$, 
where $\tilde \rho$ is the state of $A$ and some systems 
$B_n$, and $\hat \rho$ is the state of the other systems, 
then $C(\rho)=C(\tilde \rho)$, where $C$ measures the amount 
of correlations between $A$ and the considered systems $B_n$, 
since $\rho$ and $\tilde \rho$ can be transformed into each 
other by local operations. For a global state $\rho$ with genuine 
multipartite correlations, as tracing out a system $B_n$ is a local 
operation and $\xi_{d_1,d_2}$ is a nonincreasing function, 
the lowest bound on $E(\rho_A)$ is given by eq.\eqref{mineq} 
with the state $\rho$ of all the systems.

We now consider specific cases for which the boundary given 
by eq.\eqref{mineq} can be determined explicitly. A measure of 
total correlations can be defined as a mimimal distance to 
the set of product states, i.e., 
$C^{(D)}(\rho)=\inf_{\delta_A,\delta_B} 
D(\rho,\delta_A\otimes \delta_B)$, 
where the infimum is taken over all the density operators of 
$A$ and $B$, and $D$ fulfills 
$D[\Lambda(\omega),\Lambda(\omega')] \le D(\omega,\omega')$ 
for any quantum operation $\Lambda$. 
Some possible choices for $D$ are the relative entropy, 
the Bures distance $D_B$, or the Hellinger distance $D_H$ 
\cite{MPSVW,ACMBMAV,BCLA}. For the relative entropy, 
the above definition gives the mutual information \cite{MPSVW}. 
For the monotones $C^{(D_B)}$ and $C^{(D_H)}$, an explicit 
expression for $f$ can be obtained, see the Supplemental Material. 
For the entanglement of formation $E_f$, 
the entropy $s_{2,2}$ is known \cite{VADM}. Using these results, 
we find 
$$\xi_{2,2}^{E_f,D_B} (x) = u \left(x^2-\frac{x^4}{4}\right) , 
\xi_{2,2}^{E_f,D_H} (x) = u \left(\frac{x^2}{2} \right) ,$$
where $x$ varies from $0$ to $1$ for $C^{(D_B)}$, and from $0$ 
to $\sqrt{3/2}$ for $C^{(D_H)}$. The expression of $u$ is given in 
the Supplemental Material. 

Figure \ref{Fig} displays these two functions. They both vanish 
on a finite interval. As a consequence, for any state $\rho$ 
such that $C(\rho)$ exceeds a threshold value, the local 
entanglement $E(\rho_A)$ necessarily vanishes, whereas, for 
any amount of correlations $x$ below this threshold, there are 
states $\rho$ such that $C(\rho)=x$ and $\rho_A$ is entangled. 
The existence of this threshold also implies that $C(\rho)$ is at 
a finite distance from the maximum value $c_d$ as soon 
as $E(\rho_A)$ is not zero. As shown in the Supplemental Material, 
this feature is not specific to the particular cases considered above. 
Moreover, the threshold is the same for all the monotones $E$ 
vanishing only for separable states.
 
As seen above, for some correlation monotones, 
$C(\rho)=c_d$ ensures the vanishing of $E(\rho_A)$. 
On the contrary, for any monotones $C$ and $E$, 
and dimension $d_1$, there are states $\rho$ for which 
$E(\rho_A)=e_{d_1}$ and $C(\rho)$ is as high as we wish, 
provided $d_2$ is large enough. They are pure states $\rho$ 
such that the reduced density operator 
$\rho_A=\sum_i p_i |\phi_i \rangle \langle \phi_i |$ is a mixed 
maximally entangled state \cite{LZFFL}. That is to say, 
the eigenvectors of $\rho_A$ are of the form 
$|\phi_i \rangle=
\sum_{j=1}^{d_1} |j \rangle_1 |ij \rangle_2 /\sqrt{d_1}$, 
where $|j \rangle_1$ are orthonormal states of 
${\cal H}_{d_1}$, and $|ij \rangle_2$ of ${\cal H}_{d_2}$, 
i.e., ${_2\langle} ij|i'j' \rangle_2=\delta_{i,i'}\delta_{j,j'}$. 
As $\rho$ is pure, 
$C(\rho)=f({\boldsymbol p})$, and, provided $d_2/d_1$ is 
large enough, there is ${\boldsymbol p}$ such that 
$C(\rho) \ge x$, where $x$ is any amount of correlations. 
For any entanglement monotone $E$, 
$E(\rho_A)=E(|\phi_1 \rangle \langle \phi_1 |)=e_{d_1}$, 
since $\rho_A$ and $|\phi_1 \rangle \langle \phi_1 |$ can be 
transformed into each other by local operations, that do not 
affect one subsystem of $A$ \cite{fn}. Note that, though 
$E(\rho_A)=e_{d_1}$ does not imply $C(\rho)=0$ in general, 
this is true for the entanglement of formation $E_f$ and 
$d_2<2d_1$, since, for such dimensions, the only states 
$\rho_A$ for which $E_f(\rho_A)$ is maximum are pure 
\cite{LZFFL}. 

The above Theorem applies to many kinds of external 
correlations, as discussed below Lemma \ref{lem1}. 
When $C$ is an entanglement monotone, it generalizes 
previously obtained results \cite{PRL}. As mentioned above, 
$C$ can also be a measure of total correlations, or the HV 
measure of classical correlations. For this last correlation 
monotone, equation \eqref{ESloc} is an equality for 
some classical-classical states
$\rho=\sum_{i,j} p_{ij} 
|i\rangle_A{_A \langle}i| \otimes|j\rangle_B{_B \langle}j|$, 
where $|i\rangle_A$ are orthonormal states of $A$, 
$|i\rangle_B$ of $B$, and $ p_{ij}$ are probabilities summing 
to unity \cite{PHH,OHHH}. They are the strictly correlated 
classical-classical states, i.e., such that 
$p_{ij}=p_i\delta_{i,j}$ \cite{HV}. 
Consequently, there are not only pure 
states but also classical-classical states close to the boundary 
given by eq.\eqref{mineq}, for any amount of correlations. 
Moreover, since $\xi_{d_1,d_2}(c_d)=0$, 
this shows that, even when external correlations are purely 
classical, the maximum accessible local entanglement 
decreases to zero as they increase. 

In general, it can be proved that the classical-classical 
states $\rho$ obey eq.\eqref{ESloc} with $f$ replaced 
by a function $\tilde f \le f$, such that 
$C(\rho)=\tilde f[\boldsymbol \lambda (\rho_A)]$ 
when $\rho$ is strictly correlated, see the Supplemental 
Material. Provided $\tilde f$ is continuous, it follows that, 
for a classical-classical state $\rho$, $E(\rho_A)$ and 
$C(\rho)$ satisfy eq.\eqref{mineq} with $\xi_{d_1,d_2}$ 
replaced by an a priori different function $\zeta_{d_1,d_2}$. 
When $C$ is an entanglement monotone, this is 
meaningless, since $C(\rho)=0$ for all classical-classical 
states $\rho$. As seen above, for the HV measure, 
$\zeta_{d_1,d_2}=\xi_{d_1,d_2}$. For other 
correlation monotones, they obviously fulfill 
$\zeta_{d_1,d_2}\le \xi_{d_1,d_2}$. For the measure of total 
correlations $C^{(D_H)}$ and the entanglement of formation 
$E_f$, we find 
$\zeta_{2,2}^{E_f,D_H} = \xi_{2,2}^{E_f,D_B}$, 
see the Supplemental Material and Fig.\ref{Fig}. For the mutual 
information, $\tilde f$ is the Shannon entropy $h$. 
For this correlation monotone, inequality \eqref{ESloc} with 
$h$ in place of $f$, and hence 
$E(\rho_A)  \le \zeta_{d_1,d_2}[C(\rho)]$, is actually valid 
for all separable states $\rho$, as $S(\rho_B) \le S(\rho)$ for 
any separable state $\rho$ \cite{HH}, and, since 
$f=2\tilde f=2h$, $\zeta_{d_1,d_2} (x) = \xi_{d_1,d_2} (2x)$ 
where $x\in[0,\ln d]$, for any entanglement monotone $E$.

We finally discuss the relations of other local properties 
to external correlations. A first natural question is whether 
$E$ can be replaced by any correlation monotone  in 
inequality \eqref{mineq}. Lemma \ref{lem2} is not specific 
to entanglement monotones. It only requires that $E$ is 
convex \cite{PRL}. Many familiar 
entanglement monotones are convex, though this is not 
a basic requirement for such a measure \cite{HHHH}. 
For other correlation monotones, imposing convexity can 
lead to some difficulties. A convex correlation monotone is 
necessarily zero for all separable states. 
The measures of total correlations 
considered above do not vanish for all separable states, by 
construction, and are hence not convex. Consequently, 
the above derivation of eq.\eqref{mineq} does not apply if 
$E$ is replaced by anyone of these measures. Entanglement 
is not the only quantum resource for which there are measures 
that vanish only for free states and are convex. Other examples 
are the nonuniformity, which can be quantified by 
$\ln d-S(\rho_A)$ for a system $A$ of Hilbert space dimension 
$d$ \cite{GMNSH}, and the coherence, which can be quantified 
by $-\sum_i p_i \ln p_i-S(\rho_\mathrm{A})$, 
where $p_i=\langle i | \rho_\mathrm{A} | i \rangle$ and 
$\{ | i \rangle \}_i$ is the basis with respect to which 
the incoherent states are defined \cite{BCP}. In both 
these cases, inequality \eqref{mineq} is satisfied with 
the above corresponding measure in place of $E$, $\ln d-x$ 
in place of $\xi_{d_1,d_2}(x)$, and any correlation monotone 
$C$ for which $f=h$ \cite{PRL}. 
A relation of the form of eq.\eqref{mineq} can also be 
obtained for contextuality quantifiers \cite{PRA,PRA2}.

In summary, we have shown that internal entanglement and 
external correlations limit each other, whatever the nature of 
the correlations. For a given amount of external correlations 
$C(\rho)$, the internal entanglement $E(\rho_A)$ can approach 
but not exceed a value that decreases with increasing $C(\rho)$, 
and reciprocally. For familiar correlation monotones, $E(\rho_A)$ 
vanishes when the correlations are maximal. The entanglement 
can even be suppressed for lower values of $C(\rho)$. In two 
particular cases, we have determined explicitly the tight upper 
bound on $E(\rho_A)$ set by $C(\rho)$, and found that 
the entanglement vanishes when the amount of correlations is 
above a threshold value. Such a threshold also exists for other 
entanglement and correlation monotones. On the contrary, 
a maximum internal entanglement does not always ensure that 
the external correlations vanish, due to the existence of mixed 
maximally entangled states \cite{LZFFL}. If $E$ is 
the entanglement of formation, for example, this is only true 
if none of the subsystems of $A$ has a Hilbert space dimension 
larger, or equal, than twice that of the other one. As we have 
seen, the generalization of our result to other internal correlations 
is not obvious with the approach we have used. But it may be 
correct, and it would be of interest to determine whether this is 
indeed so.

\section{Supplemental Material}

In this Supplemental Material, we prove the results mentioned in 
the main text.

\subsection{Proof of Lemma 1}

Consider any probability vector $\boldsymbol p$, and any pure 
states $| \psi_1 \rangle$ and $| \psi_2 \rangle$, with Schmidt 
coefficients $\sqrt{p_i}$, of the bipartite Hilbert spaces 
${\cal H}_1 \otimes \tilde {\cal H}_1$ and 
${\cal H}_2 \otimes \tilde {\cal H}_2$, respectively. 
The Hilbert spaces ${\cal H}_1$ and ${\cal H}_2$ can always 
be considered as subspaces of a larger Hilbert space ${\cal H}$, 
and similarly 
$\tilde {\cal H}_1,\tilde {\cal H}_2 \subset \tilde {\cal H}$. 
Moreover, there are unitary operators $U$ and $\tilde U$, 
on ${\cal H}$ and $\tilde {\cal H}$, respectively, 
such that $| \psi_2 \rangle=U \otimes \tilde U | \psi_1 \rangle$. 
Thus, $| \psi_1 \rangle  \langle \psi_1 |$ and 
$| \psi_2 \rangle \langle \psi_2 |$ can be transformed into 
each other by local operations. Consequently, the amount of 
correlations $C(| \psi \rangle \langle \psi |)$ is the same for 
all the pure states $| \psi \rangle$ with Schmidt coefficients 
$\sqrt{p_i}$. We name it $f(\boldsymbol p)$. 
For $\boldsymbol p=1$, $| \psi \rangle$ is necessarily 
a product state, and so $f(1)=0$.

Consider any systems $A$ and $B'$, and any state $\rho$ 
of the composite system $AB'$, consisting of $A$ and $B'$, 
with Hilbert space ${\cal H}_{AB'}$. Denote its eigenvalues 
by $\mu_m$ and the corresponding eigenstates by 
$| m \rangle$. Let us introduce a third system, say $B''$, 
which constitutes, together with $B'$, system $B$. Provided 
the dimension of ${\cal H}_{B''}$ is not smaller than that 
of ${\cal H}_{AB'}$, $\rho$ can be written as 
$\rho=\operatorname{tr}_{B''}| \psi \rangle \langle \psi |$, 
where $\operatorname{tr}_{B''}$ is the partial trace over 
$B''$, and $| \psi \rangle=
\sum_m \sqrt{\mu_m} | m \rangle | \tilde m \rangle$ is 
a pure state of system $AB$, with orthonormal states 
$| \tilde m \rangle$ of $B''$. As $\operatorname{tr}_{B''}$ 
is a local operation, performed on $B$, 
$C(\rho) \le C(| \psi \rangle \langle \psi |)$. 
Since $\operatorname{tr}_B 
| \psi \rangle \langle \psi | = \operatorname{tr}_{B'} \rho 
= \rho_A$, the Schmidt coefficients of $| \psi \rangle$, as 
a pure state of ${\cal H}_A \otimes {\cal H}_B$, are 
$\sqrt{\lambda_i(\rho_A)}$, and hence 
$C(| \psi \rangle \langle \psi |)
=f[\boldsymbol \lambda(\rho_A)]$, which finishes the proof.

\subsection{Dependence on $d_1$ and $d_2$ of the entropy 
$s_{d_1,d_2}$}

The entropy $s_{d_1,d_2}$ is defined by
$$s_{d_1,d_2}(\boldsymbol p)=e_{d_1}-\sup_{\{ |i \rangle \}} 
E \Big( \sum_i p_i |i \rangle \langle i | \Big) , $$
where the supremum is taken over the orthonormal basis sets 
$\{ |i \rangle \}$ of ${\cal H}_{d_1} \otimes {\cal H}_{d_2}$, 
for $\boldsymbol p \in {\cal E}_{d_1d_2}$, and by 
$s_{d_1,d_2}(\boldsymbol p)=e_{d_1}$ otherwise \cite{PRL}. 
First note that $s_{d_1,d_2}$ depends on $d_1$, since 
its maximum value is $e_{d_1}$. We now show that 
the entropies $s_{2,d}$ can be different from each other even 
on ${\cal E}_4$, where they are all given by the above 
expression involving $E$. Consider an entanglement 
monotone $E$ that vanishes only for separable states, e.g., 
the entanglement of formation. In this case, for any 
$\boldsymbol p \in {\cal E}_{2d}$, 
$s_{2,d}(\boldsymbol p)=e_2$ if and only if all the states 
$\rho_A$ on ${\cal H}_{2} \otimes {\cal H}_{d}$ such that 
$\boldsymbol \lambda (\rho_A)=\boldsymbol p$ are separable. 
Consequently, for any $\boldsymbol p \in {\cal E}_{2d}$,
$s_{2,d}(\boldsymbol p)=e_2$ if and only if 
$p_1 \le p_{2d-1}+2\sqrt{p_{2d-2}p_{2d}}$, 
where $p_i=0$ for $i$ larger than the size of $\boldsymbol p$ 
\cite{J}. So, if $d \ge 3$ then $s_{2,d}<e_2$ on ${\cal E}_4$, 
whereas $s_{2,2}$ reaches $e_2$ on ${\cal E}_4$, e.g., 
for $(1/4,\ldots,1/4)$, and hence $s_{2,d} \ne s_{2,2}$ 
on ${\cal E}_4$. 

\subsection{Proof of Lemma 3}

For any $\boldsymbol p \in {\cal E}_d$, we define, 
for $\beta \ge 1$, the family of vectors 
$\boldsymbol p^{(\beta)} \in {\cal E}_d$ as follows. 
We denote by $r$ the size of $\boldsymbol p$. 
If $r=1$ or $p_2<p_1$, the components of 
$\boldsymbol p^{(\beta)}$ are given 
by $p_i^{(\beta)} \propto (p_i/p_1)^\beta$. 
Clearly, $\boldsymbol p^{(\beta)}$ is continuous with 
$\beta$, $\boldsymbol p^{(1)}=\boldsymbol p$, and 
$\boldsymbol p^{(\infty)}=1$. If there is an index 
$j>1$ such that $p_j=p_1$ and, if $j<r$, 
$p_{j+1}<p_1$, consider the vectors 
$\boldsymbol {\tilde p}^{(\eta)}$ given by 
$\tilde p_1^{(\eta)}=p_1+\eta$, 
$\tilde p_i^{(\eta)}=p_1-\eta/(j-1)$ 
for $i\in \{2, \ldots, j\}$, and, if $j<r$, 
by $\tilde p_i^{(\eta)}=p_i$ for $i > j$. There is 
$\eta^*>0$ such that, for any $\eta \in [0,\eta^*]$, 
the components of $\boldsymbol {\tilde p}^{(\eta)}$ are 
in decreasing order, and so 
$\boldsymbol {\tilde p}^{(\eta)} \in {\cal E}_d$. 
In this case, we define $\boldsymbol p^{(\beta)}$ by 
$\boldsymbol p^{(\beta)}
=\boldsymbol {\tilde p}^{(\beta-1)}$ 
for $\beta \in [1,1+\eta^*]$, and by $p_i^{(\beta)} \propto 
(\tilde p^{(\eta^*)}_i/\tilde p^{(\eta^*)}_1)^{\beta-\eta^*}$ 
for $\beta > 1+\eta^*$. Here also $\boldsymbol p^{(\beta)}$ 
is continuous with $\beta$, and 
$\boldsymbol p^{(1)}=\boldsymbol p$. Moreover, 
since $\tilde p^{(\eta^*)}_2<\tilde p^{(\eta^*)}_1$, 
$\boldsymbol p^{(\infty)}=1$. 

We denote by ${\cal E}_d(x)$ the set of all 
$\boldsymbol p \in {\cal E}_d$ such that 
$f(\boldsymbol p)=x$, and define the function $g_d$, 
on the set $I$ of the values of $x$, by $$g_d(x)=
\inf_{\boldsymbol p \in {\cal E}_d(x)} s(\boldsymbol p) .$$
By construction, $g_d \circ f \le s$ on ${\cal E}_d$, and 
there is $\boldsymbol p \in {\cal E}_d(x)$ such that 
$s(\boldsymbol p)$ and $g_d(x)$ are as close to each other 
as we wish. As $s(1)=f(1)=0$, there is a set ${\cal E}_d(0)$ 
containing $\boldsymbol p=1$, and $g_d(0)=0$. 
If $f(1/d, \ldots, 1/d)=c_d$ and $f(\boldsymbol p)<c_d$ 
for all other vectors $\boldsymbol p \in {\cal E}_d$, then 
${\cal E}_d(c_d)$ is the singleton $\{ (1/d, \ldots, 1/d) \}$, 
and hence $g_d(c_d)=s(1/d, \ldots, 1/d)$.

Due to the extreme value theorem, the continuous function 
$f$ has a maximum, $c_d$, on the simplex ${\cal E}_d$. 
Let $\boldsymbol q$ be a vector of ${\cal E}_d$ such 
that $f(\boldsymbol q)=c_d$, and define 
$\boldsymbol q^{(\beta)}$ as explained above. As $f$ is 
continuous, $f(\boldsymbol q^{(\beta)})$ is a continuous 
function of $\beta$. It is equal to $c_d$ for $\beta=1$, and 
to $0$ for $\beta \rightarrow \infty$. So, due to 
the intermediate value theorem, for any $x \in [0,c_d]$, 
there is $\hat \beta$ such that 
$f(\boldsymbol q^{(\hat \beta)})=x$. Thus, $I$ is equal to 
this interval. For any $\boldsymbol p \in {\cal E}_d(x)$, 
$f(\boldsymbol p^{(\beta)})$ is a continuous function 
of $\beta$, which is equal to $x$ for $\beta=1$, and to $0$ 
for $\beta \rightarrow \infty$. So, for any $y \in [0,x]$, 
there is $\hat \beta \ge 1$ such that 
$\boldsymbol p^{(\hat \beta)} \in {\cal E}_d(y)$. Moreover, 
since $\boldsymbol p$ is majorized by 
$\boldsymbol p^{(\hat \beta)}$ \cite{MOA,CRP}, and $s$ is 
an entropy, 
$s(\boldsymbol p)  \ge s(\boldsymbol p^{(\hat \beta)})
\ge g_d(y)$. Consequently, for any $y \le x$, $g_d(y)$ is 
a lower bound of $s$ on ${\cal E}_d(x)$, which implies 
that $g_d$ is nondecreasing.

\subsection{Expressions of $C^{(D_B)}$ and $C^{(D_H)}$ 
for specific states}

The Bures distance is given by 
$D_B(\omega,\omega')=(2-2\operatorname{tr} 
\sqrt{\sqrt{\omega} \omega'\sqrt{\omega}})^{1/2}$.
The ensuing correlation monotone $C^{(D_B)}$ reads, 
for pure states,$$C^{(D_B)}(| \psi \rangle \langle \psi |)=
\inf_{\delta_A,\delta_B} \big(2-2 \langle \psi | 
\delta_A\otimes \delta_B | \psi \rangle^{1/2} \big)^{1/2} .$$
With the Schmidt form 
$| \psi \rangle=\sum_i \sqrt{p_i} |i\rangle_A|i\rangle_B$, 
where $|i\rangle_A$ are orthonormal states of $A$, 
$|i\rangle_B$ of $B$, and the probabilities $p_i$ 
are in decreasing order, one can write 
$$\langle \psi | \delta_A\otimes \delta_B | \psi \rangle=
\sum_{i,j} \sqrt{p_i p_j} {_A\langle} i |\delta_A|j\rangle_A
{_B\langle} i |\delta_B|j\rangle_B .$$
Using this expression, the Cauchy-Schwarz inequality, 
and $p_i \le p_1$, leads to 
$$\langle \psi | \delta_A\otimes \delta_B | \psi \rangle^2 
\le p^2_1 \operatorname{tr} (\delta_A^2)
\operatorname{tr} (\delta_B^2)\le p^2_1.$$
For $\delta_A=|1\rangle_A{_A\langle} 1 |$ and 
$\delta_B=|1\rangle_B{_B\langle} 1 |$, the above inequalities 
are equalities, and hence
$$C^{(D_B)}(| \psi \rangle \langle \psi |)
=\sqrt{2(1-\sqrt{p_1})}=f^{(D_B)}(\boldsymbol p) .$$

The Hellinger distance is given by 
$D_H(\omega,\omega')=(2-2\operatorname{tr}
\sqrt{\omega} \sqrt{\omega'})^{1/2}$, and thus
$$C^{(D_H)}(| \psi \rangle \langle \psi |)=
\inf_{\delta_A,\delta_B} \big(2-2 \langle \psi | 
\sqrt{\delta_A}\otimes \sqrt{\delta_B} | 
\psi \rangle \big)^{1/2} .$$ Following the same steps 
as above, we obtain 
$$C^{(D_H)}(| \psi \rangle \langle \psi |)
=\sqrt{2(1-p_1)}=f^{(D_H)}(\boldsymbol p) .$$ 
For a strictly correlated classical-classical state 
$\rho_{pc}=\sum_{i} p_{i} 
|i\rangle_A{_A \langle}i| \otimes|i\rangle_B{_B \langle}i|$, 
with the probabilities $p_i$ in decreasing order, one finds 
$$C^{(D_H)}(\rho_{pc})=
\inf_{\delta_A,\delta_B} \Big[2-2 \sum_{i} \sqrt{p_{i}} 
\big(\sqrt{\delta_A}\big)_{ii}   \big(\sqrt{\delta_B}\big)_{ii}   
\Big]^{1/2} .$$
where $(\sqrt{\delta_{A/B}})_{ii}={_{A/B} \langle}i| 
\sqrt{\delta_{A/B}} |i\rangle_{A/B}$. 
As above, the infimum is a minimum reached for 
$\delta_A=|1\rangle_A{_A\langle} 1 |$ 
and $\delta_B=|1\rangle_B{_B\langle} 1 |$, and so
$$C^{(D_H)}(\rho_{pc})=f^{(D_B)}(\boldsymbol p)
=\tilde f^{(D_H)}(\boldsymbol p) .$$

\subsection{Derivation of $\xi_{2,2}$ for the entanglement 
of formation and the correlation monotones $C^{(D_B)}$ 
and $C^{(D_H)}$}

For the entanglement of formation $E_f$, the entropy 
$s_{2,2}$ is given by
\begin{equation}
s_{2,2}^{(E_f)}(\boldsymbol p)=\ln 2 
- v\big[\max\big\{0,p_1-p_3-2\sqrt{p_2p_4}\big\}\big] , 
\nonumber
\end{equation}
where $v(y)=w_+(y)+w_-(y)$, with $w_\pm(y)=
-(1\pm \sqrt{1-y^2})\ln [(1\pm \sqrt{1-y^2})/2]/2$ 
\cite{VADM}, and $e^{(E_f)}_2=\ln 2$. For the considered 
correlation monotones, the condition $f(\boldsymbol p)=x$ 
can be rewritten as $p_1=1-y$, where $y=x^2-x^4/4$ for 
$C^{(D_B)}$, and $y=x^2/2$ for $C^{(D_H)}$. From 
this expression and $\sum_i p_i=1$, it follows that 
the minimum value of $s_{2,2}^{(E_f)}(\boldsymbol p)$ 
under the constraint $f(\boldsymbol p)=x$, can be obtained 
by maximizing 
$z(p_2,p_4)=1-2y+(\sqrt{p_2}-\sqrt{p_4})^2$, 
since $v$ is an increasing function. The ordering of 
the probabilities $p_i$, i.e., $p_i \ge p_{i+1}$, imposes 
$\partial_{p_2} z \ge 0$, $\partial_{p_4} z \le 0$, 
$p_4 \ge 0$, $p_2 \le 1-y$, $2p_2+p_4 \ge y$, and 
$2p_4+p_2 \le y$. Three cases must be distinguished: 
$y \in [0,1/2]$, $y \in [1/2,2/3]$, and $y \in [2/3,3/4]$. 
In the first, $z$ reaches its maximum, $1-y$, for 
$(p_2,p_4)=(y,0)$. In the second, $z$ is maximum 
for $(p_2,p_4)=(1-y,0)$, where it is equal to $2-3y$. 
In the last case, $z$ is negative. The corresponding 
values for $\xi_{2,2}(x)=u(y)$ are $v(1-y)$, 
$v(2-3y)$, and $v(0)=0$. 

\subsection{Functions $\xi_{d_1,d_2}$ vanishing on a finite 
interval}

If $f=k \circ s_{R\alpha}$ where $k$ is a strictly increasing 
function and $s_{R\alpha}$ is the R\'enyi entropy of order 
$\alpha \in (0,1]$, then $\xi_{d_1,d_2}$ vanishes on 
a finite interval for any $d_1$, $d_2$, and entanglement 
monotone $E$.
\begin{proof}
Consider any vectors $\boldsymbol p$ and $\boldsymbol q$ 
of ${\cal E}_d$, where $d$ is any positive integer, and 
pad them with zeros, if necessary, to make up 
the $d$-component vectors $\boldsymbol {\tilde p}$ and 
$\boldsymbol {\tilde q}$. The R\'enyi divergence $D_\alpha$ 
of order $\alpha \in (0,1]$ is related to the total variation 
distance $V(\boldsymbol {\tilde p}, \boldsymbol {\tilde q})
=\sum_i |{\tilde p}_i- {\tilde q}_i|$, 
by the generalized Pinsker's inequality 
$D_\alpha(\boldsymbol {\tilde p}||\boldsymbol {\tilde q}) \ge 
\alpha V(\boldsymbol {\tilde p}, \boldsymbol {\tilde q})^2/2$ 
\cite{vEH,G}, and hence 
$D_\alpha(\boldsymbol {\tilde p}||\boldsymbol {\tilde q}) \ge 
\alpha \sum_i ({\tilde p}_i- {\tilde q}_i)^2/2$. 
Since $D_\alpha[\boldsymbol {\tilde p}||(1/d, \ldots, 1/d)]
=\ln d-s_{R\alpha}(\boldsymbol p)$, and $k$ is increasing, 
the function $f=k\circ s_{R\alpha}$ satisfies 
$f(\boldsymbol p) \le k(\ln d-\alpha[P(\boldsymbol p)-1/d]/2)$, 
where $P(\boldsymbol p)=\sum_i p_i^2$. 

For any $d_1 \ge 2$ and $d_2 \ge d_1$, all the states 
$\rho_A$ on ${\cal H}_{d_1} \otimes {\cal H}_{d_2}$ for 
which $P[\boldsymbol \lambda (\rho_A)] \le 1/(d-1)$, 
where $d=d_1d_2$, are separable, and hence such that 
$E(\rho_A)=0$ for any entanglement monotone $E$ 
\cite{ZHSL,GB}. Thus, due to Lemma 2, for any 
$\boldsymbol p \in {\cal E}_d$ such that 
$P(\boldsymbol p) \le 1/(d-1)$, 
$s_{d_1,d_2}(\boldsymbol p)=e_{d_1}$. Consequently, 
as $k$ is strictly increasing, for any 
$\boldsymbol p \in {\cal E}_d(x)$ with 
$x \ge k[\ln d-\alpha/2d(d-1)]$, 
$s(\boldsymbol p)=e_{d_1}$, where $s=s_{d_1,d_2}$, 
and thus, for any such $x$, $g_d(x)=e_{d_1}$, see 
the definition of $g_d$. In other words, 
$\xi_{d_1,d_2}=e_{d_1}-g_d$ vanishes on a finite interval. 
\end{proof}
Consider two entanglement monotones $E$ and $E'$ 
which are zero only for separable states. 
If $\xi^{(E,C)}_{d_1,d_2}$ corresponding to $E$ 
and the correlation monotone $C$, vanishes on 
an interval $J$ and is positive elsewhere, then 
$\xi^{(E',C)}_{d_1,d_2}$ vanishes on $J$ and is positive 
elsewhere.
\begin{proof}
With the entropy $s_{d_1,d_2}$ ($s'_{d_1,d_2}$) given 
by Lemma 2 with $E$ ($E'$), and $f$ given by Lemma 1 
with $C$, define the function $g_d$ ($g'_d$), with 
$d=d_1d_2$, as above. Denote by $J$ the maximal 
interval on which $\xi^{(E,C)}_{d_1,d_2}=e_{d_1}-g_d$ 
vanishes. For any $\boldsymbol p \in {\cal E}_d(x)$ 
with $x \in J$, it follows from the definition of $g_d$ 
and from the fact that $s_{d_1,d_2}$ cannot exceed 
$e_{d_1}$, that 
$s_{d_1,d_2}(\boldsymbol p)=e_{d_1}$. Thus, 
due to Lemma 2 and the assumption on $E$, all 
the density operators $\rho_A$ on 
${\cal H}_{d_1} \otimes {\cal H}_{d_2}$ of 
spectrum $\boldsymbol \lambda(\rho_A)=\boldsymbol p$ 
are separable, and so such that $E'(\rho_A)=0$, which 
gives $s'_{d_1,d_2}(\boldsymbol p)=e'_{d_1}$. 
Consequently, $J$ is a subset of $J'$ the maximal interval 
on which $\xi^{(E',C)}_{d_1,d_2}=e'_{d_1}-g'_d$ 
vanishes. Switching the roles of $E$ and $E'$ in 
the above arguments leads to $J'=J$.
\end{proof}

\subsection{Boundary for the classical-classical states}

For any correlation monotone $C$, there is a function 
$\tilde f$ of the probability vectors with $\tilde f(1)=0$, 
such that, for any classical-classical state $\rho$, 
$C(\rho) \le \tilde f[\boldsymbol \lambda (\rho_A)]$, 
with equality when $\rho$ is strictly correlated.

Let ${\cal H}_{d_1} \otimes {\cal H}_{d_2}$, 
with $d_2 \ge d_1$, be the Hilbert space of system 
$A$, and $d=d_1d_2$. For a convex entanglement 
monotone $E$, and a correlation monotone $C$ 
such that $\tilde f$ is continuous, $C(\rho)$ and 
$E(\rho_A)$ obey, for any classical-classical state 
$\rho$, $E(\rho_A)  \le \zeta_{d_1,d_2}[C(\rho)] $, 
where $\zeta_{d_1,d_2}$ is a nonincreasing function 
on $[0,\tilde c_d]$, with $\tilde c_d$ the maximum 
of $\tilde f$ on ${\cal E}_d$, such that 
$\zeta_{d_1,d_2}(0)=e_{d_1}$. 
For any $x \in [0,\tilde c_d]$, there are classical-classical 
states $\rho$ such that $C(\rho)=x$, and the two sides 
of the above inequality are as close to each other as 
we wish. 
 
If $\tilde f(1/d, \ldots, 1/d)=\tilde c_d$ and 
$\tilde f(\boldsymbol p)<\tilde c_d$ for any other 
$\boldsymbol p \in {\cal E}_d$, then 
$\zeta_{d_1,d_2}(\tilde c_d)=0$. 
\begin{proof}
Consider any probability vector $\boldsymbol p$, and 
any strictly correlated classical-classical states 
$\rho_1$ and $\rho_2$, of the bipartite Hilbert spaces 
${\cal H}_1 \otimes \tilde {\cal H}_1$ and 
${\cal H}_2 \otimes \tilde {\cal H}_2$, respectively, 
such that $\rho_k
=\sum_i p_i |i\rangle_k {_k\langle} i | \otimes 
|\tilde \imath \rangle_k {_k\langle} \tilde \imath |$, 
where $|i\rangle_k$ are orthonormal states of 
${\cal H}_k$, and $|\tilde \imath \rangle_k$ 
of $\tilde {\cal H}_k$. The Hilbert spaces ${\cal H}_1$ 
and ${\cal H}_2$ can always be considered as 
subspaces of a larger Hilbert space ${\cal H}$, 
and similarly 
$\tilde {\cal H}_1,\tilde {\cal H}_2 \subset \tilde {\cal H}$. 
Moreover, there are unitary operators $U$ and $\tilde U$, 
on ${\cal H}$ and $\tilde {\cal H}$, respectively, such that 
$\rho_2=U \otimes \tilde U \rho_1 
U^\dag \otimes \tilde U^\dag$. Thus, the amount of 
correlations $C(\rho)$ is the same for all the strictly 
correlated classical-classical states $\rho$ with 
eigenvalues $p_i$. We name it $\tilde f(\boldsymbol p)$. 
For $\boldsymbol p=1$, $\rho$ is necessarily a product 
state, and so $\tilde f(1)=0$.

Consider any systems $A$ and $B$, and any 
classical-classical state 
$\rho=\sum_{i,j} p_{ij} 
|i\rangle_A{_A \langle}i| \otimes|j\rangle_B{_B \langle}j|$, 
where $\{ |i\rangle_A \}_i$ is an orthonormal basis of $A$, 
$\{ |i\rangle_B \}_i$ of $B$, and $ p_{ij}$ are probabilities 
summing to unity. The corresponding reduced density 
operator for $A$ reads 
$\rho_A=\sum_i p_i |i\rangle_A{_A \langle}i|$, 
where $p_i=\sum_j p_{ij}$. The local operation with Kraus 
operators 
$K_{ij}=\sqrt{p_{ij}/p_i} I \otimes |j\rangle_B{_B \langle}i|$, 
where $I$ is the identity operator of $A$, for $i$ such that 
$p_i \ne 0$ and any $j$, and 
$K_i=I \otimes |i\rangle_B{_B \langle}i|$ for $i$ such that 
$p_i=0$, changes the strictly correlated classical-classical 
state $\hat \rho=\sum_i p_i |i\rangle_A{_A \langle}i| 
\otimes|i\rangle_B{_B \langle}i|$ into $\rho$, and so 
$C(\rho) \le C(\hat \rho)$. The state of $A$ is $\rho_A$ 
for both $\rho$ and $\hat \rho$. From 
$\boldsymbol \lambda (\rho_{A})=\boldsymbol p$, 
it follows that 
$C(\hat \rho)=\tilde f[\boldsymbol \lambda (\rho_{A})]$. 
Using then Lemma 2 and Lemma 3 with $\tilde f$ and 
the entropy $s_{d_1,d_2}$ finishes the proof.
\end{proof}


\begin{thebibliography}{99}
\bibitem{fn2} For any $\rho$, $C(\tilde \rho)=C(\rho)$ for all 
$\tilde \rho=U\rho U^\dag$, where $U$ is any unitary
operator of $A$, whereas there is such a $U$ changing $\rho_A$ 
into the separable state 
$\tilde \rho_A = \sum_i \lambda_i(\rho_A) |i\rangle \langle i|$, 
where $|i\rangle$ are product states of 
${\cal H}_{d_1} \otimes {\cal H}_{d_2}$, for which 
$E(\tilde \rho_A)=0$. For any $\rho_A$, $C(\rho)=0$ for all 
$\rho=\rho_A\otimes \rho_B$.
\bibitem{fn} The operation with Kraus operators 
$\sqrt{p_i} I \otimes U_i$, where $I$ is the identity operator 
on ${\cal H}_{d_1}$, and $U_i$ a unitary operator on 
${\cal H}_{d_2}$ such that $|i j \rangle_2=U_i|1j \rangle_2$, 
transforms $|\phi_1 \rangle \langle \phi_1 |$ into $\rho_A$. 
A basis of ${\cal H}_{d_2}$ can always be formed with 
the states $|ij \rangle_2$ and, if necessary, other ones, denoted 
$|l \rangle_2$. The operation with Kraus operators 
$I \otimes \sum_{j=1}^{d_1} |1j \rangle_2 {_2\langle} ij|$, 
and, if necessary, $I \otimes |l \rangle_2 {_2\langle} l|$, 
changes $\rho_A$ into $|\phi_1 \rangle \langle \phi_1 |$.
\end{thebibliography}
\end{document}